# Substrate interaction mediated control of phase separation in FIB milled Ag-Cu thin films

Vivek C. Peddiraju, Pravallika Bandaru, Shourya Dutta-Gupta, Subhradeep Chatterjee*

*Department of Materials Science and Metallurgical Engineering, Indian Institute of Technology Hyderabad, Telangana - 502284, India*

*Email:* subhradeep@msme.iith.ac.in

**Abstract**

Nanofabrication is an integral part of realization of advanced functional devices ranging from optical displays to memory devices. Focused ion beam (FIB) milling is one of the widely used nanofabrication methods. Conventionally, FIB milling has been carried out for patterning single-phase stable thin films. However, the influence of FIB milling on phase separation of metastable alloy films during subsequent treatments has not been reported. Here, we show how FIB milling of Ag-Cu thin films influences the separation process and microstructure formation during post-milling annealing. Phase-separated microstructure of the film consists of fine, randomly distributed Ag-rich and Cu-rich domains, whereas adjacent to milled apertures (cylindrical holes), we observe two distinctly coarser rings. A combination of imaging and analysis techniques reveals Cu-rich islands dispersed in Ag-rich domains in the first ring next to the aperture, while the second ring constitutes mostly of Ag-rich grains. Copper silicide is observed to form in and around apertures through reaction with the Si-substrate. This substrate interaction, in addition to known variables like composition, temperature, and capillarity, appears to be a key element in drastically changing the local microstructure around apertures. This current study introduces new avenues to locally modulate the composition and microstructure through an appropriate choice of the film-substrate system. Such an ability can be exploited further to tune device functionalities with possible applications in plasmonics, catalysis, microelectronics and magnetics.

## I. Introduction

Nanostructuring of *single-phase* thin films using masks or focused ion beam (FIB) milling is often employed to fabricate devices with enhanced or novel properties[1–8]. For phase-separated films, however, controlling a pattern during milling can be challenging due to the difference in sputtering rates of the individual components[9]. As an alternative, nanostructures can be fabricated in the metastable single-phase films of immiscible systems, which can then be subjected to an annealing heat treatment to induce phase separation. Therefore, it is of great importance, both at practical and fundamental levels, to investigate how phase separation is affected by the pre-existing nanostructuring present in the film.

Phase-separation phenomena in Ag-Cu bare thin films (that is, without imprinted nanostructuring) have been studied extensively[10–14]. Control of the separation process can potentially be exploited to tune the optical response of these films[9,15]. Alloy films produced by physical vapor deposition usually exist in a metastable single-phase, solid solution state[9–11], although initiation of phase separation process during deposition has also been reported in some cases[16–19]. When as-deposited films are subjected to a heat treatment, Ag- and Cu-rich domains form via phase separation of the undecomposed or partly decomposed film by nucleation and growth or spinodal decomposition[11,14,15,20]. Surface energies of the components (Ag and Cu in this case), interfacial energy, and lattice mismatch between them as well as with the substrate are known to influence the decomposition process[21–30]. Here we report unusual phase separation patterns observed around cylindrical holes (termed as 'apertures') fabricated by FIB milling on sputter-deposited Ag-Cu films. Further experiments and analysis demonstrate substrate-film interaction to be responsible for the striking shift in the decomposition process in the film.

## II. Materials and Methods

*A. Thin film deposition*: Ag-Cu thin films are deposited on a substrate by magnetron co-sputtering with pure Ag and Cu targets (99.99% purity) under an Ar (99.99 % pure) atmosphere. Single crystal p-type Si-wafers of ~380 μm thickness with a (100) orientation and 20 nm thick 100 $\mu$m × 100 $\mu$m amorphous $SiN_x$ (ASN) TEM windows were used as substrates. Deposition parameters (70 W RF and 30 W DC power for Ag and Cu, respectively; initial chamber pressure: $9 \times 10^{-7}$ mbar and pressure during deposition: $4.4 \times 10^{-3}$ mbar; Ar flow rate: 5 SCCM; deposition time: 60 s) are optimized to deposit approximately equimolar (~53 ± 1 at.% Ag), ~38 nm thick films.

*B. FIB milling and post-treatment*: Apertures of different diameters on as-deposited films are fabricated by FIB milling. Further experimental details of the milling are provided as Supplementary Information (SI); Table S1 lists the nominal and measured aperture sizes. Multiple replicas are made for each aperture diameter to ascertain the reproducibility of the results. Films with apertures are annealed at 400 °C for 30 minutes under vacuum to allow for phase separation.

*C. Characterization*: Composition of the films is measured by energy dispersive X-ray spectroscopy (EDS; make & model: EDAX Octane Elect Plus) in a multi-beam (electron plus FIB) scanning electron microscope (SEM; make & model: JEOL JIB 4700F) operated at 20 kV. The same SEM (at 15 and 20 kV) is also used for imaging; unless stated otherwise, composition sensitive backscattered electron (BSE) mode is used for imaging. Atomic force microscopy (AFM of make Park Systems) of film cross section is used to measure film thickness which is cross verified by SEM imaging. Further microstructural analysis is done in a 200 kV scanning transmission electron microscope (S/TEM; make & model: JEOL JEM F-200) equipped with a high angle annular dark field (HAADF) detector. Unless explicitly mentioned otherwise, all SEM images are from films on Si, and TEM/STEM images and electron diffraction patterns are from those deposited on ASN windows. SEM and TEM images are analysed further to reveal local phase and compositional details. Steps in image processing and representative examples are provided in SI. Characterization experiments are carried out after each of the individual steps, *viz.*, deposition, FIB milling and annealing.

## III. Results and Discussion

The experimental workflow is illustrated by the schematics in Fig. 1(a). The first step involves co-deposition of Ag and Cu which is same for both the substrates (schematic in the left). This is followed by FIB milling to fabricate apertures: for the Si substrate, these extend slightly into the substrate (schematic in the middle), whereas through-holes are made in films on ASN windows. SE and BSE images of an FIB cut section of an aperture for the film on Si are provided Fig. S1.

Fig. 1(b) shows a BSE image of an as-deposited film with an aperture of ~400 nm diameter. A TEM diffraction ring pattern obtained from a similar film, along with the corresponding rotationally averaged intensity profile as a function of radial distance are shown in Fig 1(b). The intensity profile reveals the presence of two distinct peaks (111 and 200) that overlap to from the broad, brightest ring in pattern.

These peaks fit to a single-phase metastable solid solution of Ag and Cu. The lattice parameter of this as-deposited FCC phase is 3.91 Å. This is in between the lattice parameters of pure Ag and Cu, indicating an extended solid solubility of Ag and Cu. The uniform contrast in the BSE image along with this diffraction evidence confirm the lack of phase separation at this scale of observation in the as-deposited film. Additional evidence of compositional homogeneity in as-deposited film is presented in the STEM image and EDS maps of Fig. S1(d).

Upon annealing, films undergo phase separation, as evidenced by domains of two distinct mean gray levels in the HAADF-STEM image of Fig. 1(c). The corresponding STEM EDS elemental map reveals the bright and dark domains to be Ag- and Cu-rich, respectively. Additionally, the electron diffraction pattern of this region clearly shows the occurrence of distinct rings that belong to the two different FCC phases (Ag-rich and Cu-rich). Lattice parameters of these two FCC phases are 4.07 Å and 3.62 Å, which are close to lattice parameters of pure Ag and Cu, respectively. Thus, all the evidence confirm that phase separation has taken place in the metastable as-deposited film during annealing.

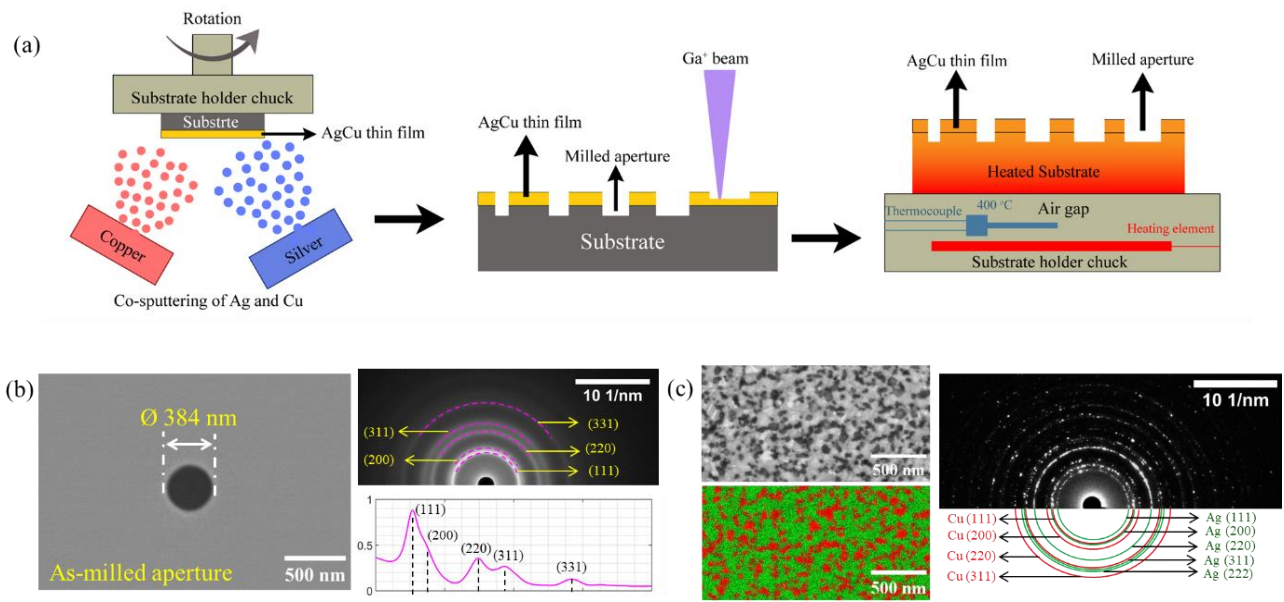

*Fig. 1. (a) Schematic of the workflow. Left: Deposition of equimolar Ag-Cu alloy films using DC magnetron sputtering. Middle: Fabrication of apertures on as-deposited films using FIB milling. Right: Vacuum annealing of films with nanostructures inside the deposition chamber. (b) Left: Representative SEM-BSE image of an aperture milled on as-deposited film (Si substrate). Right: Electron diffraction (ED) ring pattern of the as-deposited film (ASN substrate) along with the corresponding rotationally averaged intensity profile. (c) HAADF-STEM image*

*(top left) of the annealed film (ASN substrate) along with a corresponding STEM-EDS elemental map (bottom left) (Cu: red, Ag: green). Annotated ED pattern on the right show rings from Ag-rich and Cu-rich phases.*

BSE image of Fig. 2(a) shows the post-annealing microstructure of a region containing a 400 nm aperture. Here, the bright blocky particle at the centre obscures the milled aperture; a secondary electron image shown as an inset reveals the actual aperture being covered by this particle. Strikingly, microstructure in the vicinity of the aperture is very different from that far away from it (see Fig. 1(c)). Adjacent to the aperture, we observe two distinct and approximately annular regions which are marked by dashed lines and indicated as $A_1$ and $A_2$. A magnified view of these zones is presented in Fig. 2(b). Zone $A_1$ consists of a mixture of Ag- and Cu-rich domains whereas $A_2$ contains mostly Ag-rich grains.

Fig. 2(c) shows the EDS spectra obtained from different microstructural features that are present in Fig 2(a). The Si peak at 1.3 KeV is not included in the range as strong substrate contribution obscures other peaks. Spectrum from the bright particle in the center shows a very strong Cu-K signal in comparison to Cu-K signal from other microstructural features and careful quantitative analysis estimates its stoichiometry close to $Cu_3Si$. The Cu-enrichment in and around the aperture is also evident from the EDS composition maps presented in Fig. 2(d). These results suggest the formation of a copper silicide phase in the central region. Additional evidence of formation of $Cu_3Si$ is presented through the cross-sectional image in Fig. S2(a). It shows the particle to extend into the Si-substrate with a sharp V-shaped interface that is characteristic of copper silicide particles forming on Si-(100) substrate[31–34]. Although this particle is enriched in Cu (lower atomic number), the relatively bright BSE image contrast originates from its surface topography (projecting out of the plane), as confirmed by AFM measurements in Fig S2(b) and Fig. S2(c).

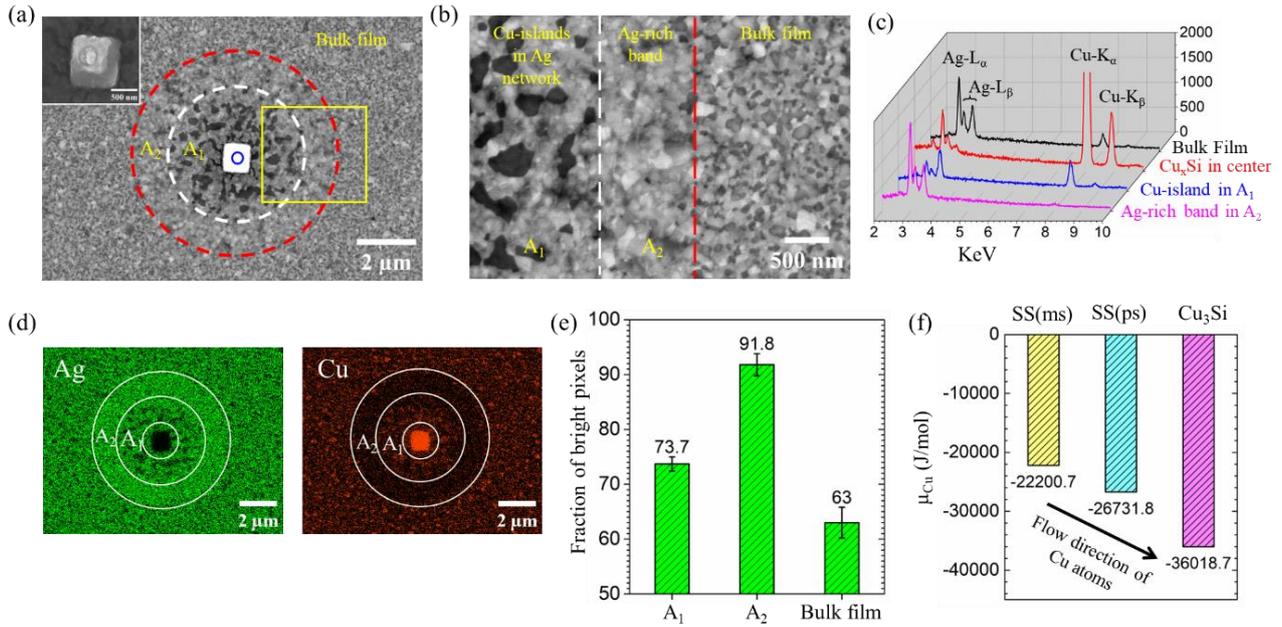

*Fig. 2. Phases and microstructure of the annealed film with an aperture (Si substrate). (a) Low magnification BSE image showing formation of halo structure (red dashed circle) around the aperture (blue circle). The white dashed circle demarcates $A_1$ and $A_2$ zones within the halo. Secondary electron (SE) image in the inset shows a magnified view of copper silicide (bright particle at center). (b) Magnified view of region marked by yellow box in (a). (c) Stack of SEM-EDS spectra obtained from different microstructural features shown in (a). (d) SEM-EDS elemental maps highlight the overall redistribution of Ag and Cu in the halo region. (e) Semi-quantitative information about Ag concentration obtained through fraction of bright pixels in different microstructural regions. (f) Bar chart showing chemical potentials of copper in different phases, viz., metastable solid solution (SS(ms)), phase separated mixture (SS(ps)), and $Cu_{19}Si_6$ (nominally $Cu_3Si$). Arrow denotes the direction of transport of Cu atoms from SS(ms) towards $Cu_3Si$.*

Formation of the distinct microstructural zones $A_1$ and $A_2$ around the aperture needs to be discussed further. We confirmed (Fig. S3) that they form regardless of the aperture size and their spatial extent is constant across the aperture sizes. The dark islands in $A_1$ are Cu-rich (~90 at. % Cu) while the bright interconnected network is rich in Ag (~70 at. % Ag). Adjacent to $A_1$, the circular banded region $A_2$ appears with a relatively bright contrast compared to rest of the film. It is mostly Ag-rich (~90 at. % Ag), although a few darker grains (possibly Cu-rich) scattered among the numerous Ag-rich band could also be observed in Fig. 2(b). Due to the limited spatial resolution of the SEM-EDS, it is not possible to capture such fine-scale composition modulations. The microstructure within the two annular bands $A_1$ and $A_2$ is called as 'halo structure'. We note a gradual transition from the region $A_1$ to region $A_2$, but transition from $A_2$ to the bulk microstructure is sharp with a discernible boundary between them. To understand the redistribution of Ag and Cu within the halo structure, we carry out image analysis (procedure outlined in Supplementary Information) and represent the Ag-enrichment of each ring using the fraction of bright pixels in Fig. 2(e). This clearly establishes the depletion of Cu from regions surrounding the aperture.

As stated earlier, we observe the formation of a copper silicide phase inside the milled aperture after annealing. Hong *et al.* had reported [35] the formation of $Cu_3Si$ by interface reaction between an Ag-Cu thin film with Si substrate. Therefore, presence of this compound phase must be considered while trying to understand the phase separation behavior of the undecomposed Ag-Cu solid solution. To understand the thermodynamics of the influence of substrate interaction on phase separation in the metastable film, chemical potentials of Cu ($\mu_{Cu}$) in the metastable solution (SS(ms)), phase separated Cu-rich solution (SS(ps)) and $Cu_{19}Si_6$ phase (nominally $Cu_3Si$) need to be compared. We comput these potentials using the computational thermodynamics software Thermo-Calc® with the SSOL6 database.

The bar plot in Fig. 2(f) shows that $\mu_{Cu}^{Cu_{19}Si_6} < \mu_{Cu}^{SS(ps)} < \mu_{Cu}^{SS(ms)}$. Therefore, transformation during the annealing is initiated by a reaction between the exposed Si substrate inside the aperture and the undecomposed film to form $Cu_3Si$. This depletes the aperture-adjacent regions of Cu and sets up a flux of Cu atoms from the bulk film towards the aperture and modulates the phase separation process in the film taking place by spinodal decomposition. Fig. 2(e) provides evidence of the increase in the Ag-content in the annular regions $A_1$ and $A_2$ due to the depletion of Cu from these regions. Surprisingly, however, we find the Cu-depletion to be greater in $A_2$ than $A_1$. Interaction between compound formation inside the aperture and spinodal decomposition within the Ag-Cu film likely results in such unusual microstructural and compositional patterns. Two alternative growth scenarios are presented in Fig. 3 to explain their origin.

Fig. 3(a) depicts a case where the tendency to form $Cu_3Si$ in the aperture creates a gradient of Cu concentration towards the aperture and a corresponding gradient of Ag in the opposite direction. Thus, as $Cu_3Si$ starts to form in the aperture, it simultaneously creates alternate Cu- and Ag-rich rings due to these radial composition gradients. Eventually, phase separation in the bulk of the film too initiates and it creates finer compositional domains. Formation of these rings around the aperture would be facilitated by different surface energies of Ag and Cu[23,36]. Although this turn of events is plausible, we do not observe a uniformly Cu-rich ring around the aperture.

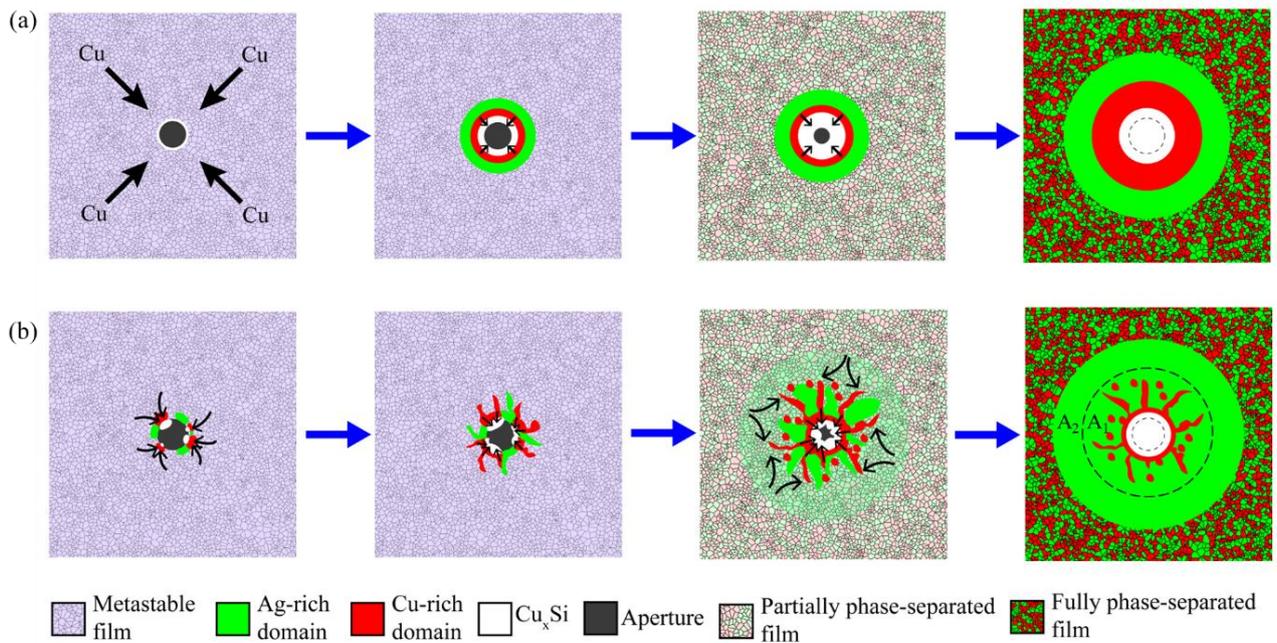

*Fig. 3. Schematic depicting two alternative scenarios of microstructural evolution (progression of time indicated by arrows); see text for details. (a) A core-shell structure around the aperture resulting from an entirely radial species transport. (b) A halo structure as observed in the experiments results when the radial symmetry is broken by discrete silicide nucleation events at the aperture-film interface. Arrows inside the schematics indicate the flux of copper atoms.*

Instead, the alternative scenario shown in Fig. 3(b) appears more likely. Here too there is an initial flux of Cu atoms is created by the formation of $Cu_3Si$ in the aperture. However, it nucleates and grows from certain locations (and does not cover the entire exposed film-aperture interface), thus breaking the radial symmetry of the flux of Cu atoms. This leads to both radial and lateral atom transport of Cu and Ag around the aperture, creating Cu-rich domains dispersed among Ag-rich ones (zone $A_1$). However, due to the loss of Cu atoms from the film to the aperture (as is required for the formation of $Cu_3Si$), there still remains a net radial inward flow of Cu atoms (and corresponding radial outward flow of Ag), creating an Ag-rich annular region $A_2$ following the mixed Ag-Cu domains in $A_1$. Eventually, kinetics of spinodal decomposition in the bulk film becomes appreciable and it creates much finer compositional domains. Thus, the microstructural pattern adjacent to the aperture gets drastically modified due to the interaction of the chemical reaction forming $Cu_3Si$ and creating the gradient in Cu.

The effect of substrate interaction (or lack thereof) is demonstrated in Fig. 4(a) for films deposited on ASN window. We can observe a slight Ag-enrichment of a narrow rim surrounding the aperture. This is supported by image analysis results in Fig. 4(b) (represented as before by the fraction of bright pixel in the annular region) and the STEM-EDS maps of Fig. 4(c). However, formation of $Cu_3Si$ and distinct annular regions surrounding the aperture are conspicuously absent for ASN. Similar observations are made for other aperture sizes too (Fig. S4). For Si substrate, Si atoms are readily available for the growth of $Cu_3Si$, whereas for the ASN substrate, the Si-N bonds must first be broken to form silicides. Studies show[37–39] that neither Ag nor Cu atoms react with $Si_3N_4$ and from compound phases at elevated temperatures. The slight enrichment of Ag around the aperture that is observed for the ASN window can be attributed to its lower surface energy compared to Cu which aids the segregation of Ag[23,40,41]. This provides strong evidence that the halo structure for the Si substrate is associated with the growth of $Cu_3Si$ and therefore, emphasizes the crucial role played by the substrate on the phase separation pattern around apertures.

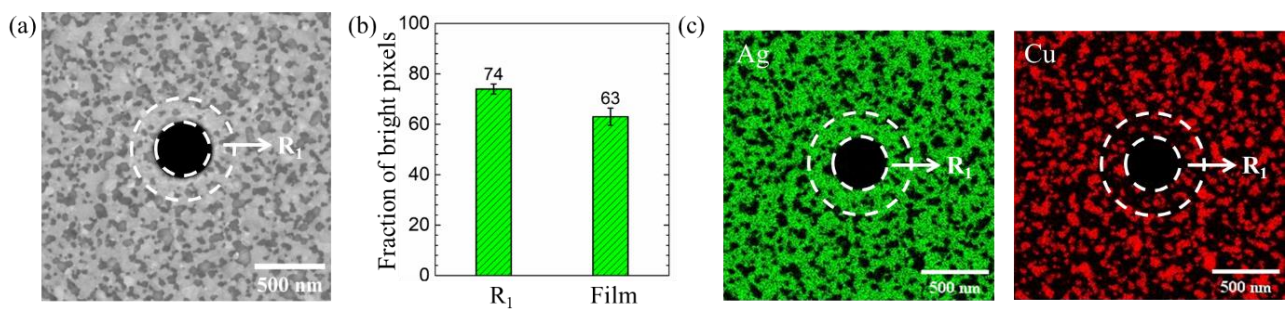

*Fig. 4. Phases and microstructure of the annealed film with an aperture (ASN substrate). (a) HAADF STEM image showing the absence of significant microstructural changes and silicides around the aperture (unlike the case of film deposited on Si substrate). (b, c) A slight Ag-enrichment around the aperture (indicated by white dashed circles in (a)) is confirmed by the bar chart of bright pixel fraction in (b) and STEM EDS maps in (c); green and red colors in the latter represent Ag and Cu concentrations, respectively.*

Results presented above demonstrate that it is possible to locally tune the multi-phase microstructures in alloy films via substrate-film reaction which in turn could give rise to enhanced properties (e.g., different plasmonic response than bulk as-deposited or phase separated film). In addition to Ag-Cu alloy thin films, the proposed method for controlling the phase separation via substrate chemical reaction can be readily extended to other phase separating and ordering alloys systems like Ag-Co, Ag-Fe, Au-Fe, Co-Cu, Co-Pt and Fe-Cu. These would enable an additional handle for controlling the response of functional devices that would be relevant for magnetic[42,43], magneto-optical[5,44–48] and catalytic[49] applications. Moreover, by introducing an additional reactive layer on the substrate (which can be inert), it is also possible to regulate the extent of such chemical reactions. In such a case, the reactive film thickness would control the availability of the reacting species and thereby provide an additional means for controlling the microstructure.

## IV. Conclusions

In summary, our results show that reaction with exposed substrate produced by FIB milled apertures influence phase separation and microstructure of Ag-Cu thin films. During the annealing treatment the growth of $Cu_3Si$ takes place inside milled apertures. Adjacent to the aperture, an annular region ($A_1$) is formed where the Cu-islands dispersed in an Ag-rich matrix. This annular region is in turn surrounded by a mostly Ag-rich band ($A_2$). The $A_1$ and $A_2$ regions together is termed as the 'halo structure'. Outside this halo, the bulk film microstructure comprises of much finer Ag-rich and Cu-rich domains that are interspersed with each other. These characteristic patterns are absent for films deposited on the ASN window where only a very thin Ag-rich annular region is surrounding the apertures is observed. The evidence presented highlights how substrate interaction with spinodal decomposition in the film can strongly modulate the local microstructure around FIB milled apertures. Therefore, results of the present study offer new insights to tailor local microstructure for enhancing material properties and device performance.

## Supplementary materials

See supplementary materials for details about experimental methods and techniques viz., thin film deposition, FIB milling, characterization, image analysis methodology and supporting microscopic images.

## Acknowledgements

The authors would like to thank Dr. Saswata Bhattacharyya and Dr. Sai Rama Krishna Malladi for fruitful discussions. S.D.-G. would like to acknowledge the funding from SERB (Grant No. SERB/ECR/2018/002628).

## Author declarations

### Conflict of interest

All authors declare that they have no conflicts to disclose.

### Author contributions

**Vivek C. Peddiraju:** Data curation (lead), Formal analysis (equal), Investigation (lead), Methodology (equal), Software (supporting), Validation (lead), Visualization (equal), Writing-Original Draft Preparation (lead) **Pravallika Bandaru:** Investigation (supporting), Writing-Review & Editing (supporting) **Shourya Dutta-Gupta:** Conceptualization (lead), Funding Acquisition (lead), Investigation (supporting), Methodology (equal), Project Administration (equal), Resources (equal), Supervision (equal), Visualization (equal), Writing-Review & Editing (equal) **Subhradeep Chatterjee:** Conceptualization (lead), Formal analysis (equal), Methodology (equal), Project Administration (equal), Resources (equal), Software (lead), Supervision (equal), Visualization (equal), Writing-Review & Editing (equal)

## Data Availability

The data that support the findings of this study are available from the corresponding author upon reasonable request.

# Substrate interaction mediated control of phase separation in FIB milled Ag-Cu thin films

Vivek C. Peddiraju, Pravallika Bandaru, Shourya Dutta-Gupta, Subhradeep Chatterjee*

*Department of Materials Science and Metallurgical Engineering, Indian Institute of Technology Hyderabad, Telangana-502284, India*

*Email: subhradeep@msme.iith.ac.in

**1. Fabrication of apertures:**

Focused ion beam milling (FIBM) technique is used to make apertures on as-deposited films. A JEOL make JIB 4700F multi-beam instrument equipped with Ga-ion source operated at 30 KV is used for this purpose. The specimen was tilted to 53° such that the $Ga^+$ probe is incident normal to the surface of the film. During the milling process, only intended (circular) region of the film was exposed to $Ga^+$ ions and rest of the film within field of view (FOV) was not exposed. This was achieved by freezing the scan at a specific magnification (thus fixing the FOV), and then physically moving the specimen stage in X and Y directions. Subsequently, Ga+ probe is incident inside the desired region (circle in this case) within this FOV thus sputtering out the material which produced the final nanostructure. In all cases, circular apertures were milled at the central location within the FOV. By systematically displacing the specimen stage (X and Y), an array of apertures with different diameters were produced. Two sets of apertures were produced using FIB milling. One set of apertures were used exclusively for imaging (SEM/TEM) and EDS analysis in the as-deposited condition. The second set of apertures were used for imaging and EDS analysis after annealing treatment. Diameters of the milled aperture (both nominal and measured) are listed in Table S1 below.

**Table S1: Diameters (in nm) of apertures fabricated on as-deposited film via FIB milling**

| Nominal diameter | Actual diameter on Si-substrate | Actual diameter on ASN substrate |
|---|---|---|
| 200 | 196 | 204 |
| 400 | 384 | 410 |
| 600 | 582 | 610 |
| 800 | 780 | 815 |
| 1000 | 978 | 1021 |

**2. Supplementary images:**

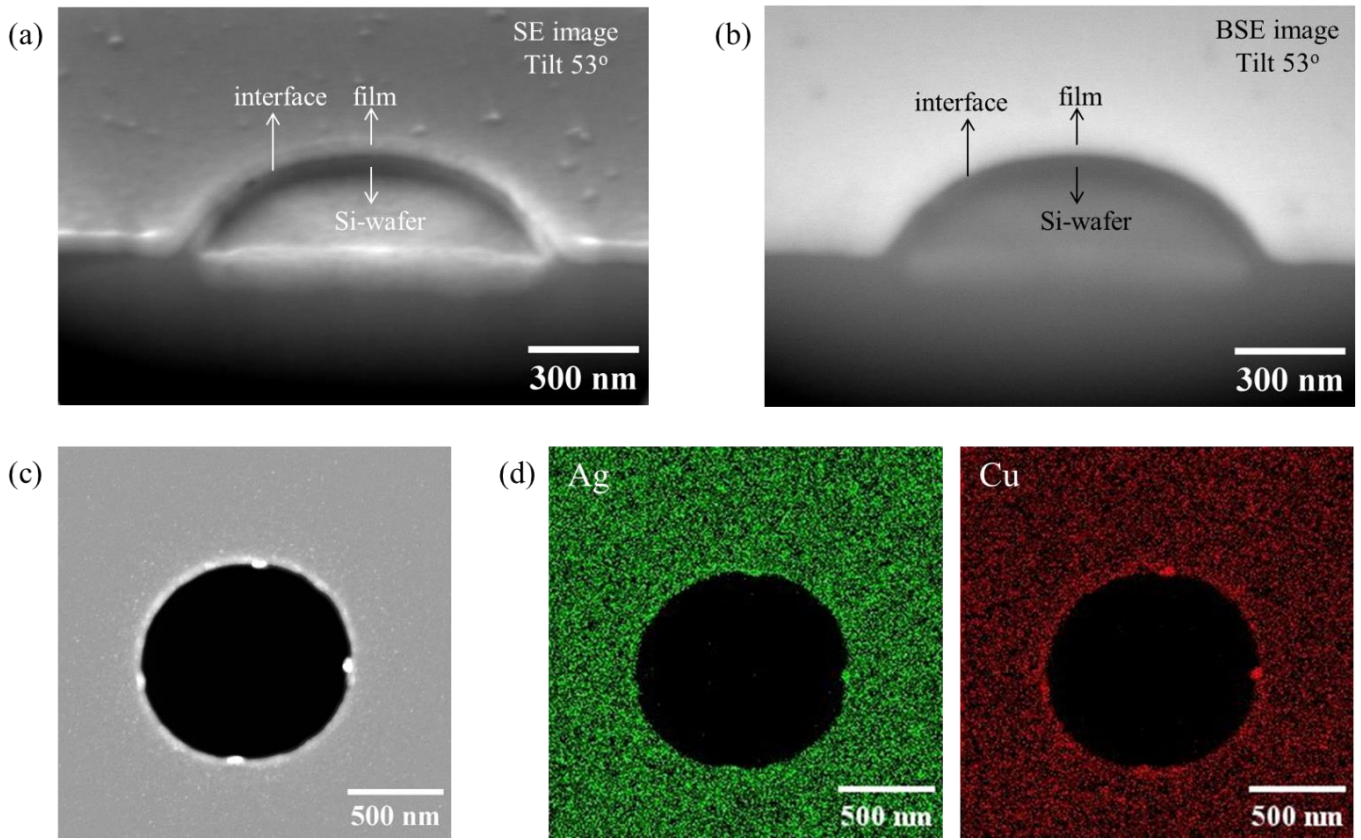

*Fig. S1. (a, b) Cross-section cut of as-milled aperture (1000 nm) imaged at 53° tilt clearly shows an aperture extending into the Si-substrate. (c) HAADF-STEM image of ~1000 nm aperture made on the ASN substate after depositing the film. (d) STEM-EDS elemental map shows uniform distribution of Ag and Cu in as-deposited condition.*

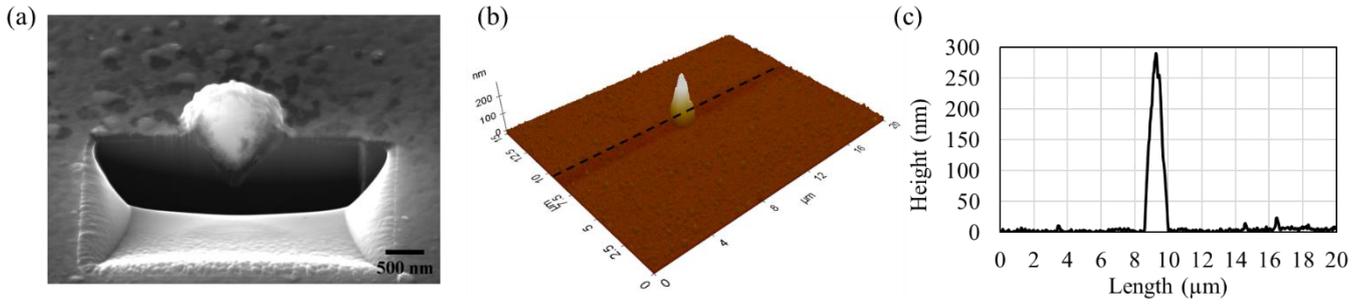

*Fig. S2. (a) SE image showing the cross-section view of the $Cu_3Si$ particle along the depth direction. (b) AFM 3D topographic image from the region shown in Figure2(a) illustrating the growth of $Cu_3Si$ normal to the surface of the film. (c) Line profile along the black-dashed line shows that the silicide phase projects ∼300 nm out of the film.*

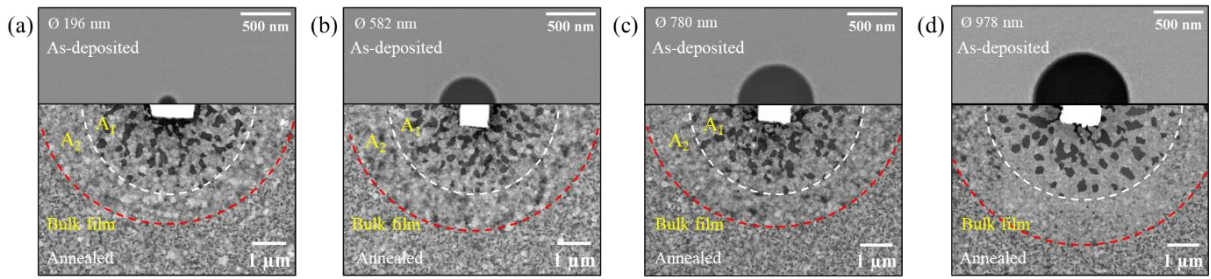

*Fig. S3. (a – d) The SEM-BSE micrographs of the as-milled apertures on films deposited on Si substrate, and the annealed films in the vicinity of the apertures. Formation of the 'halo structure' surrounding the bright copper silicide phase can be observed in all cases.*

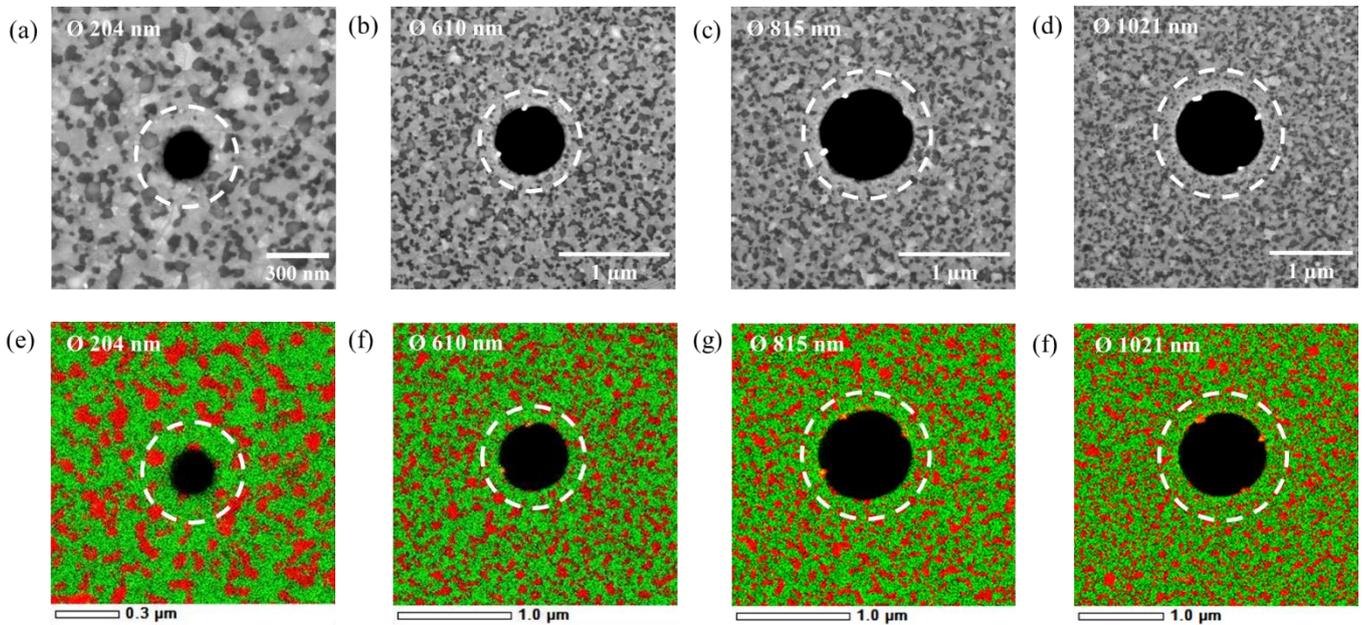

*Fig. S4. (a – d) HAADF-STEM images showing the microstructure of the film deposited on ASN substrate in the vicinity of the apertures after annealing treatment and (e-f) corresponding EDS maps. Ag is represented in green, and Cu is in red. In all the cases we notice the Ag-enrichment near the periphery of apertures.*

## 3. Image analysis:

Image analysis is carried out to identify the fraction of bright pixels from both the annual regions, $A_1$ and $A_2$, and the bulk of film, as shown in Fig. S5. The fraction of bright pixels is a direct measure of fraction of Ag-rich domains which in turn is proportional to the concentration of Ag within these regions. Firstly, the contrast of SEM-BSE image is adjusted such that it lies between 0 to 1. The image is then divided into three concentric annular segments constituting the $A_1$, $A_2$ and bulk film regions that are centred around the $Cu_xSi$. Corresponding to each annular segment, we have extracted the brightness histogram and then calculated the fraction of bright pixels within that region by identifying appropriate threshold value. A similar procedure is carried out in case of STEM images as well.

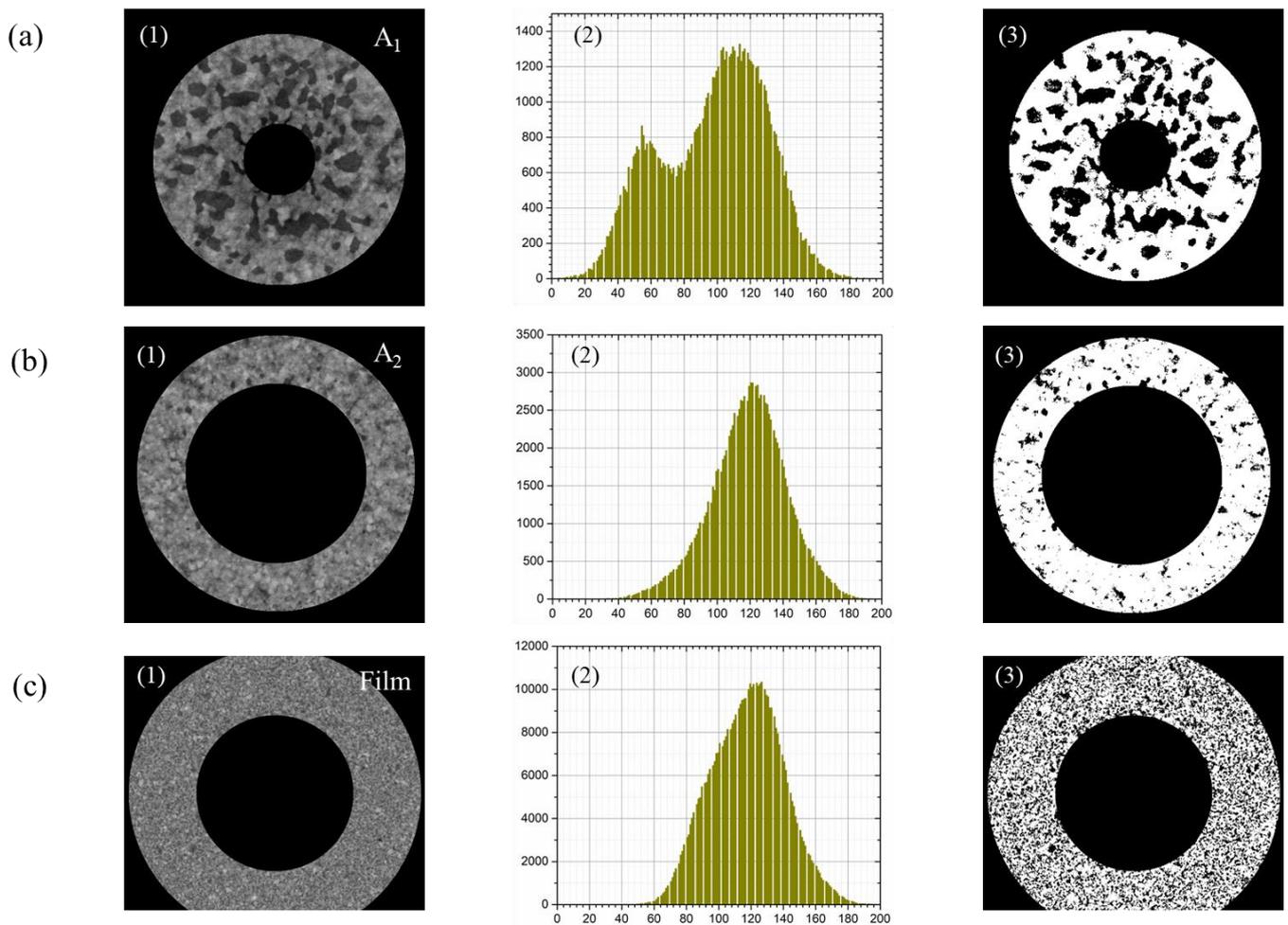

Fig. S5. (a, b, c) Image analysis of SEM micrographs corresponding to annular segments showing the region $A_1$, $A_2$ and bulk film respectively. In all three sub-figures the original SEM images are shown in (1). The image contrast histograms obtained from these regions are shown in (2). The gray scale values are along X-axis and counts

*corresponding each gray value is along Y-axis. (3) shows the binary images corresponding to the images in (1). The pixels coloured in white shows the Ag-rich region while the pixels shaded in black show the Cu-rich regions. From this thresholding, we calculate the fraction of bright contrast region within each annular segment.*